\documentstyle[preprint,eqsecnum,aps,prb]{revtex}
\renewcommand{\thesection}{\arabic{section}}
\begin{document}
\draft
\title{\Large\bf PERFECT FLUID AND TEST PARTICLE WITH SPIN AND DILATONIC
                 CHARGE IN A WEYL--CARTAN SPACE}
\author{O. V. Babourova\thanks{E-mail:baburova.physics@mpgu.msk.su}
        and B. N. Frolov\thanks{E-mail:frolovbn.physics@mpgu.msk.su}
}
\address{Department of Mathematics, Moscow State Pedagogical University,\\
     Krasnoprudnaya 14, Moscow 107140, Russia}
\maketitle
\begin{abstract}
     The equation of perfect dilaton-spin fluid motion in  the  form  of
generalized hydrodynamic  Euler-type  equation  in  a  Weyl--Cartan space is
derived. The equation of motion of a test particle with spin and
dilatonic charge in the Weyl--Cartan geometry background is obtained.  The
peculiarities of test particle motion in a Weyl--Cartan space are
discussed.
\end{abstract}
\pacs{PACS Nos: 04.20.Fy, 04.40.+c, 04.50.+h}
\newpage

\section{Introduction}
\renewcommand{\thesection}{\arabic{section}}
\markright{Perfect fluid and test particle with spin and dilatonic charge in
          a Weyl--Cartan space}
     In a  previous  work\cite{dil}  (which  we  shall  refer  as   I)   the
variational theory of the perfect fluid with intrinsic spin and
dilatonic charge (dilaton-spin fluid) was developed and the  equations  of
motion of  this  fluid,  the Weyssenhoff-type evolution equation of the spin
tensor and the conservation law of the dilatonic charge were derived.
\par
     The purpose of the present work is to investigate the equations of
motion of  such type of fluid and their consequences,  one of which leads to
the equation of motion of a test particle with spin and dilatonic charge
in the Weyl--Cartan background.
\par
     It is  well  known  that the equations of charge particle motion in
an electromagnetic theory are the consequence of the covariant
energy-momentum conservation  law  of  the system `particles--field' and the
electromagnetic field equations.\cite{LL} In General  Relativity  the
equations of matter motion are the consequence of the gravitational
field equations. The reason consists in the fact that the Einstein
equations lead to the covariant energy-momentum conservation law of matter.
In the Einstein--Cartan theory\cite{Kib}${}^{,}$\cite{Tr1} the same situation
occurs, cite{He:let} but the conservation laws have more complicated
form established  in  Ref.\onlinecite{Tr1}.  In   Refs.\onlinecite{Bab:Pon},
\onlinecite{Bab:dr} it  was  proved  that  in  the  generalized  theories of
gravity with torsion in a Riemann--Cartan space $U_{4}$ based  on  non-linear
Lagrangians the  equations  of the matter motion are also the consequence of
the gravitational field equations.  The similar result was established in  a
metric-affine space with curvature, torsion and
nonmetricity.\cite{Lan}${}^{-}$\cite{BKF2}
\par
     In Sec. 2 we shall use this method for deriving the equation of
dilaton-spin fluid motion in the form of generalized hydrodynamic Euler-type
equation in  a Weyl--Cartan space.  In Sec.  3 this equation will be applied
for obtaining the equation of motion of a test particle with  spin and
dilatonic charge in the  Weyl--Cartan geometry background.

\section{The hydrodynamic equation of motion \newline
         of the perfect dilaton-spin fluid}
\renewcommand{\thesection}{\arabic{section}}
\markright{Perfect fluid and test particle with spin and dilatonic charge in
           a Weyl--Cartan space}
\setcounter{equation}{0}
      In a Weyl--Cartan space $Y_{4}$ the matter  Lagrangian  obeys the
diffeomorfism invariance,  the  local Lorentz invariance and the local scale
invariance that leads to the corresponding  Noether identities which
can be  obtained  as  the  particular  case  of the corresponding identities
stated in a general metric-affine space\cite{He:pr}  (see  Appendix  on  the
notations used),
\begin{eqnarray}
&&{\cal D}\Sigma_{\sigma} = (\bar{e}_{\sigma}\rfloor {\cal T}^{\alpha})\wedge
\Sigma_{\alpha} - (\bar{e}_{\sigma}\rfloor {\cal R}^{\alpha}\!_{\beta})\wedge
{\cal J}^{\beta}\!_{\alpha} - \frac{1}{8}(\bar{e}_{\sigma}\rfloor {\cal Q})
\sigma^{\alpha}\!_{\alpha}\; ,  \label{eq:zak1} \\
&&\left ({\cal D} +\frac{1}{4}{\cal Q}\right )\wedge {\cal S}_{\alpha\beta} =
\theta_{ [\alpha} \wedge \Sigma_{\beta ]} \; , \label{eq:39} \\
&&{\cal D}{\cal J} = \theta^{\alpha} \wedge \Sigma_{\alpha} -
\sigma^{\alpha}\!_{\alpha}\; . \label{eq:391}
\end{eqnarray}
Here $\Sigma_{\sigma}$ is the canonical energy-momentum 3-form,
$\sigma_{\alpha\beta}$ is the metric stress-energy 4-form,
${\cal S}_{\alpha\beta}$ is the spin momentum 3-form
and ${\cal J}$ is the  dilaton  current  3-form.
\par
      In  case  of  the  perfect
dilaton-spin fluid  the  corresponding expressions for the quantities
$\Sigma_{\sigma}$, $\sigma_{\alpha\beta}$, ${\cal S}_{\alpha\beta}$ and
${\cal J}$  were  derived  in  I  (see  (I.5.3),  (I.5.5),  (I.5.6)).  These
expressions  are  compatible in the sense that they satisfy to the Noether
identities (\ref{eq:zak1}), (\ref{eq:39}) and (\ref{eq:391}). The identities
(\ref{eq:39}) and  (\ref{eq:391})  can  be  verified  with  the  help of
the spin  tensor  evolution  equation  (I.4.4)  and  the  dilatonic   charge
conservation law (I.4.2).
\par
     The Noether identity (\ref{eq:zak1}) represents  the  quasiconservation
law for  the  canonical  matter  energy-momentum  3-form.  This  identity is
fulfilled, if the equations  of  matter  motion  are  valid,  and  therefore
represents in  its essence another form of the matter motion equations.
\par
     Let us  introduce  with  the  help of (I.5.4) a specific (per particle)
dynamical momentum of a fluid element,
\begin{equation}
\pi_{\sigma}\eta := -\frac{1}{nc^{2}}*\!u\wedge \Sigma_{\sigma}\;,
\quad \pi_{\sigma} = \frac{\varepsilon}{nc^{2}}u_{\sigma} -
\frac{1}{c^{2}}S_{\sigma\rho}\bar{u}\rfloor {\cal D}u^{\rho}\; .
\label{eq:pi} \end{equation}
Then the canonical energy-momentum 3-form (I.5.4) reads,
\begin{equation}
\Sigma_{\sigma} =  p\eta_{\sigma} + n \left (\pi_{\sigma} +
\frac{p}{nc^{2}}u_{\sigma}\right ) u\; . \label{eq:sig}
\end{equation}
\par
     Substituting (\ref{eq:sig}), (I.5.5) and (I.5.6) into
(\ref{eq:zak1}), one obtains after some algebra the equation of motion of
the perfect dilaton-spin fluid in the form of the generalized hydrodynamic
Euler-type equation of the perfect fluid,
\begin{eqnarray}
&& u\wedge {\cal D} \left (\pi_{\sigma} + \frac{p}{nc^{2}}u_{\sigma}\right )
= \frac{1}{n}\eta\bar{e}_{\sigma}\rfloor {\cal D}p
- \frac{1}{8n}\eta (\varepsilon + p) Q_{\sigma} \nonumber \\
&& - (\bar{e}_{\sigma}\rfloor{\cal T}^{\alpha}) \wedge \left (\pi_{\alpha}
+ \frac{p}{nc^{2}}u_{\alpha}\right) u
- \frac{1}{2} (\bar{e}_{\sigma}\rfloor {\cal R}^{\alpha\beta})\wedge
S_{\alpha\beta}u + \frac{1}{8} (\bar{e}_{\sigma}\rfloor {\cal R}^{\alpha}\!_
{\alpha})\wedge Ju \; . \label{eq:euler}
\end{eqnarray}
\par
     Let us evaluate the component of the  equation  (\ref{eq:euler})  along
the 4-velocity  by contracting one with $u^{\sigma}$.  After some algebra we
get the energy conservation law along a streamline of the fluid,
\begin{equation}
d\varepsilon = \frac{\varepsilon + p}{n} dn \; . \label{eq:cons}
\end{equation}
Comparing this equation with the first thermodynamic principle (I.2.14),
one can conclude that along a streamline of the fluid the conditions
\begin{equation}
ds =0\;, \qquad \frac{\partial \varepsilon}{\partial S^{p}\!_{q}}
dS^{p}\!_{q} = 0\;, \qquad \frac{\partial \varepsilon}{\partial J} dJ = 0\;.
\label{eq:iso}
\end{equation}
are valid. The first of these equalities means that the entropy conservation
law is fulfilled along a streamline of the fluid. This fact corresponds to
the basic postulates of the theory.

\section{The equation of test particle motion \newline
         in a Weyl--Cartan space}
\renewcommand{\thesection}{\arabic{section}}
\markright{Perfect fluid and test particle with spin and dilatonic charge in
          a Weyl--Cartan space}
\setcounter{equation}{0}
     Let us consider the limiting case when the pressure $p$ vanishes,  then
the equation (\ref{eq:euler}) will describe the motion of one fluid particle
with the mass $m_{0} = \varepsilon/(n c^{2}) = const$, with the spin tensor
$S_{\alpha\beta}$ and the dilatonic charge  $J$,
\begin{equation}
u\wedge {\cal D} \pi_{\sigma} = - \frac{1}{8}\eta m_{0}c^{2} Q_{\sigma}
- (\bar{e}_{\sigma}\rfloor {\cal T}^{\alpha})\wedge \pi_{\alpha} u
- \frac{1}{2} (\bar{e}_{\sigma}\rfloor {\cal R}^{\alpha\beta})\wedge
S_{\alpha\beta}u + \frac{1}{8} (\bar{e}_{\sigma}\rfloor {\cal R}^{\alpha}\!_
{\alpha})\wedge Ju\; . \label{eq:test}
\end{equation}
The third term on the right-hand side of (\ref{eq:test})  represents  the
well-known Mathisson  force,  the  second  term represents the translational
force that appears in spaces with torsion.  The forth term appears only in a
Weyl--Cartan space.  It  has the Lorentz-like form with the Weyl's homothetic
curvature tensor ${\cal R}^{\alpha}\!_{\alpha}$ as dilatonic field strength.
\par
     The following Theorem is valid.
\begin{quote}
{\bf Theorem.} In a Weyl--Cartan space $Y_{4}$ the motion of a test particle
with spin and dilatonic charge obeys the
equation,
\begin{eqnarray}
&& m_{0}u\wedge \stackrel{R}{\cal D}u_{\sigma} =
\frac{1}{c^{2}}u\wedge \left (\delta^{\alpha}_{\sigma}\stackrel{C}{\cal D} -
\bar{e}_{\sigma}\rfloor {\cal T}^{\alpha}\right ) * u^{\beta}\stackrel{C}
{\cal D}(S_{\alpha\beta} u)  \nonumber \\
&& - \frac{1}{2} (\bar{e}_{\sigma}\rfloor\stackrel{C}{\cal R}\!^{\alpha\beta})
\wedge S_{\alpha\beta}u + \frac{1}{16} (\bar{e}_{\sigma}\rfloor d{\cal Q})
\wedge Ju \; , \label{eq:test1}
\end{eqnarray}
where $\stackrel{C}{\cal R}\!^{\alpha\beta}$ is a Riemann--Cartan curvature
2-form, $\stackrel{R}{{\cal D}}$  and  $\stackrel{C}{{\cal D}}$ are the
exterior covariant  differentials  with  respect  to  a Riemann connection
1-form $\stackrel{R}{\Gamma}\!^{\alpha}\!_{\beta}$ and a Riemann--Cartan
connection 1-form $\stackrel{C}{\Gamma}\!^{\alpha}\!_{\beta}$, respectively.
\end{quote}
{\bf Proof.} Using  the  decomposition  (\ref{eq:def})  (see  Appendix)  the
specific dynamical  momentum of a fluid element (\ref{eq:pi}) can be written
in the form,
\begin{equation}
\pi_{\sigma} = m_{0}u_{\sigma} - \frac{1}{c^{2}}S_{\sigma\rho}\bar{u}
\rfloor\stackrel{C}{\cal D}u^{\rho} - \frac{1}{8} S_{\sigma\rho}Q^{\rho}
\; . \label{eq:pi1}\end{equation}
With the help of the decomposition (\ref{eq:def}) one can prove that the
evolution equation of the spin tensor (I.4.4) is also valid with respect  to
the Riemann--Cartan  connection  $\stackrel{C}{\Gamma}\!^{\alpha}\!_{\beta}$
and reads,
\begin{equation}
\Pi^{\alpha}_{\sigma}\Pi^{\rho}_{\beta}\,\bar{u}\rfloor\stackrel{C}
{\cal D}S^{\sigma}\!_{\rho} = 0 \;, \label{eq:35}
\end{equation}
where $\Pi^{\alpha}_{\sigma} := \delta^{\alpha}_{\sigma} + c^{-2} u^{\alpha}
u_{\sigma}$ is the projection tensor.  Using (\ref{eq:pi}) and (\ref{eq:35})
the left-hand side of the equation (\ref{eq:test}) can be represented as
follows,
\begin{eqnarray}
&& u\wedge {\cal D} \pi_{\sigma} = m_{0}u\wedge \stackrel{C}{\cal D}
u_{\sigma} - \frac{1}{8}\eta m_{0}c^{2} Q_{\sigma}
- \frac{1}{c^{2}}u\wedge \stackrel{C}{\cal D}
(S_{\sigma\rho}\bar{u}\rfloor\stackrel{C}{\cal D}u^{\rho}) \nonumber \\
&& - \frac{1}{8}S_{\sigma\rho}u\wedge \stackrel{C}{\cal D}Q^{\rho}
- \frac{1}{64}\eta S_{\sigma\rho}Q^{\rho}Q_{\lambda}u^{\lambda}\; .
\label{eq:left}\end{eqnarray}
With the  help  of  the  decomposition  (\ref{eq:red1}) the third term on the
right-hand side of (\ref{eq:test}) takes the form,
\begin{eqnarray}
&& \frac{1}{2} (\bar{e}_{\sigma}\rfloor {\cal R}^{\alpha\beta})\wedge
S_{\alpha\beta}u = \frac{1}{2} (\bar{e}_{\sigma}\rfloor \stackrel{C}
{\cal R}\!^{\alpha\beta})\wedge S_{\alpha\beta}u
+ \frac{1}{8} \eta T^{\alpha}\!_{\sigma\rho} u^{\rho}S_{\alpha\beta} Q^{\beta}
\nonumber \\
&& - \frac{1}{8}\eta S_{\sigma\rho}\bar{u}\rfloor\stackrel{C}{\cal D}
Q^{\rho} + \frac{1}{64}\eta S_{\sigma\rho} Q^{\rho} Q_{\lambda} u^{\lambda}
\;. \label{eq:rt}
\end{eqnarray}
Substituting (\ref{eq:rt}) and (\ref{eq:pi1})  in  the  right-hand  side  of
(\ref{eq:test}), comparing the result with (\ref{eq:left}) and  taking  into
account the equality (\ref{eq:tran}),
one obtains (\ref{eq:test1}), as was to be proved.
\par
     The Theorem proved has the important consequences.
\newline
{\bf Corollary 1.} The motion of a test particle without spin and dilatonic
charge in a Weyl--Cartan space $Y_{4}$ coincides  with the motion of this
particle in the Riemann space, the metric tensor of  which  coincides  with
the metric tensor of $Y_{4}$.
\newline
{\bf Corollary  2.}  The  motion  of a test particle with spin and dilatonic
charge in a Weyl--Cartan space $Y_{4}$ coincides with the motion of this
particle in the Riemann--Cartan space, the metric tensor and the torsion
tensor of which coincide with the metric tensor and the torsion tensor of
$Y_{4}$, if one of the conditions is fulfilled:
\par i) the dilatonic field is a closed form, $\quad d{\cal Q} = 0$;
\par ii) the dilatonic charge of the particle vanishes, $\quad J = 0$.
\newline
{\bf Corollary 3.} The manifestations of the non-trivial Weyl space structure
(when the dilatonic field ${\cal Q}$ is not a closed form) can  be  detected
only with the help of the test particle endowed with dilatonic charge.
\par
    The result of the Corollary 1 can be cosidered as a particular  case  of
the Theorem  stated  in  Refs.\onlinecite{Lan} --\onlinecite{BKF2}  for the
matter motion in a general metric-affine space.

\section{Conclusions}
\renewcommand{\thesection}{\arabic{section}}
\markright{Perfect fluid and test particle with spin and dilatonic charge in
            a Weyl--Cartan space}
    The perfect dilaton-spin fluid model represents the medium with spin and
dilatonic charge which  generates  the  spacetime  Weyl--Cartan  geometrical
structure and interacts with it.  The influence of the Weyl--Cartan geometry
on dilaton-spin fluid motion is described by the Euler-type hydrodynamic
equation. This hydrodynamic equation leads to the equation of motion of a
test particle with spin and dilatonic charge in the Weyl--Cartan geometry
background, the special form of which is stated by the Theorem of Sec.~3.
\par
     The important consequences of this Theorem  mean  that  bodies  and
mediums without  dilatonic  charge are not subjected to the influence of the
possible Weyl structure of spacetime (in contrast to the generally  accepted
opinion) and  therefore  can  not  be  the  tools  for  the detection of the
Weyl properties of spacetime. For `usual' matter without dilatonic charge
the Weyl structure of spacetime is unobservable. In order to investigate the
different manifestations of the possible Weyl structure of spacetime one
needs to use the bodies and mediums endowed with dilatonic charge.

\appendix
\section*{}
\renewcommand{\theequation}{\thesection.\arabic{equation}}
\markright{Perfect fluid and test particle with spin and dilatonic charge in
          a Weyl--Cartan space}
\setcounter{equation}{0}
     Let us consider a connected 4-dimensional oriented differentiable
manifold ${\cal M}$ equipped with a linear connection $\Gamma$ and a metric
$g$ of index 1, which are not compatible in general in the sense that
the covariant  exterior differential of the metric does not vanish,
\begin{equation}
{\cal D}g_{\alpha\beta} = dg_{\alpha\beta} - \Gamma^{\gamma}\!_{\alpha}
g_{\gamma\beta} - \Gamma^{\gamma}\!_{\beta}g_{\alpha\gamma}
=: -{\cal Q}_{\alpha\beta}\; , \label{eq:301}
\end{equation}
where $\Gamma^{\alpha}\!_{\beta}$ is a connection 1-form and ${\cal Q}_
{\alpha\beta}$ is a nonmetricity 1-form, ${\cal Q}_{\alpha \beta } =
Q_{\alpha \beta \gamma} \theta^{\gamma}$.
\par
     A curvature 2-form ${\cal R}^{\alpha}\!_{\beta}$ and a torsion 2-form
${\cal T}^{\alpha}$,
\begin{equation}
{\cal R}^{\alpha}\!_{\beta}=\frac{1}{2}R^{\alpha}\!_{\beta\gamma\lambda}
\theta^{\gamma}\wedge\theta^{\lambda}\;,
\qquad {\cal T}^{\alpha}=\frac{1}{2}T^{\alpha}\!_{\beta\gamma}\theta^{\beta}
\wedge\theta^{\gamma}\;,
\label{eq:302}
\end{equation}
are defined by virtue of the Cartan's structure equations,
\begin{eqnarray}
{\cal R}^{\alpha}\!_{\beta}=d\Gamma^{\alpha}\!_{\beta}+\Gamma^{\alpha}\!_
{\gamma}\wedge\Gamma^{\gamma}\!_{\beta}\;, \qquad
{\cal T}^{\alpha} = {\cal D}\theta^{\alpha} = d\theta^{\alpha}+\Gamma^
{\alpha}\!_{\beta}\wedge\theta^{\beta}\;. \label{eq:304}
\end{eqnarray}
\par
     A Weyl--Cartan space $Y_{4}$ is a space with a metric,  curvature,
torsion and nonmetricity which obeys the constraint (${\cal Q}$ is a Weyl
1-form),
\begin{equation}
{\cal Q}_{\alpha\beta} = \frac{1}{4}g_{\alpha\beta}{\cal Q}\;, \qquad
{\cal Q}:= g^{\alpha\beta}{\cal Q}_{\alpha\beta} = Q_{\alpha}\theta^{\alpha}
\;. \label{eq:371}
\end{equation}
\par
     In a Weyl--Cartan space the following decomposition of the connection
is valid,
\begin{equation}
\Gamma^{\alpha}\!_{\beta} = \stackrel{C}{\Gamma}\!^{\alpha}\!_{\beta} +
\Delta^{\alpha}\!_{\beta}\; , \qquad
\Delta^{\alpha}\!_{\beta} = \frac{1}{8}(2\theta^{[\alpha}Q_{\beta ]} +
\delta^{\alpha}_{\beta} {\cal Q})\; , \label{eq:def}
\end{equation}
where $\stackrel{C}{\Gamma}\!^{\alpha}\!_{\beta}$   denotes  a  connection
1-form of a Riemann--Cartan space $U_{4}$ with curvature, torsion and
a metric compatible with a connection.
\par
     The decomposition (\ref{eq:def}) of the connection induces corresponding
decomposition of the curvature,
\begin{eqnarray}
&&{\cal R}^{\alpha}\!_{\beta} = \stackrel{C}{{\cal R}}\!^{\alpha}\!_{\beta} +
\stackrel{C}{{\cal D}}\Delta^{\alpha}\!_{\beta}
+ \Delta^{\alpha}\!_{\gamma}\wedge \Delta^{\gamma}\!_{\beta} =
\stackrel{C}{{\cal R}}\!^{\alpha}\!_{\beta} + {\cal P}^{\alpha}\!_{\beta}
+ \frac{1}{8}\delta^{\alpha}_{\beta}d{\cal Q}\;, \quad
{\cal P}^{\alpha}\!_{\beta} = {\cal P}^{[\alpha}\!_{\beta ]}\; ,
\label{eq:red1} \\
&&{\cal P}^{\alpha}\!_{\beta} =
\frac{1}{4}\left ({\cal T}^{ [\alpha}Q_{\beta ]} - \theta^{ [\alpha}\wedge
\stackrel{C}{\cal D}Q_{\beta ]} + \frac{1}{8}\theta^{ [\alpha}Q_{\beta ]}
\wedge {\cal Q} - \frac{1}{16} \theta^{\alpha} \wedge \theta_{\beta}
Q_{\gamma}Q^{\gamma}\right )\; , \label{eq:red2}
\end{eqnarray}
where $\stackrel{C}{{\cal D}}$ is the exterior covariant differential with
respect to the Riemann--Cartan connection 1-form $\stackrel{C}{\Gamma}\!^
{\alpha}\!_{\beta}$ and $\stackrel{C}{{\cal R}}\!^{\alpha}\!_{\beta}$ is the
Riemann--Cartan curvature 2-form.  In (\ref{eq:red1}) the last term  contains
the Weyl homothetic curvature 2-form,
\begin{equation}
{\cal R}^{\alpha}\!_{\alpha} = \frac{1}{2}{\cal D}{\cal Q} =
\frac{1}{2}(\bar{e}_{\alpha}\rfloor {\cal D}Q_{\beta})\theta^{\alpha}\wedge
\theta^{\beta} + \frac{1}{2}Q_{\alpha}{\cal T}^{\alpha} =
\frac{1}{2}d{\cal Q}\;. \label{eq:seg}
\end{equation}
\par
     The Riemann--Cartan connection 1-form can be decomposed as follows,
\begin{eqnarray}
&&\stackrel{C}{\Gamma}\!^{\alpha}\!_{\beta} = \stackrel{R}{\Gamma}\!^
{\alpha}\!_{\beta} + {\cal K}^{\alpha}\!_{\beta}\;, \quad
{\cal T}^{\alpha} =: {\cal K}^{\alpha}\!_{\beta}\wedge \theta^{\beta}\;,
\label{kon1}\\
&& {\cal K}_{\alpha\beta} = 2\bar{e}_{[\alpha}\rfloor {\cal T}_{\beta ]} -
\frac{1}{2} \bar{e}_{\alpha}\rfloor \bar{e}_{\beta}\rfloor ({\cal T}_{\gamma}
\wedge \theta^{\gamma})\;, \label{kon2}
\end{eqnarray}
where $\stackrel{R}{\Gamma}\!^{\alpha}\!_{\beta}$ is a Riemann
(Levi--Civita) connection 1-form  and  ${\cal  K}^{\alpha}\!_{\beta}$  is  a
kontorsion 1-form.\cite{He:pr} In a Riemann--Cartan  space  the covariant
differentiation with respect to the transport connection\cite{Tr2} is useful,
\begin{equation}
\delta_{\sigma}^{\rho}\stackrel{tr}{{\cal D}} := \delta_{\sigma}^{\rho}
\stackrel{C}{{\cal D}} - \bar{e}_{\sigma}\rfloor {\cal T}^{\rho}\; .
\end{equation}
In particular, the following equality is valid,
\begin{equation}
u\wedge \stackrel{tr}{{\cal D}}u_{\sigma} =
u\wedge \left (\delta^{\rho}_{\sigma}\stackrel{C}{{\cal D}} -
\bar{e}_{\sigma}\rfloor {\cal T}^{\rho}\right ) u_{\rho} =
u\wedge \stackrel{R}{{\cal D}} u_{\sigma}\;.  \label{eq:tran}
\end{equation}

\end{document}